\documentstyle{article}
\begin{document}
\title{Cryptanalysis of Key Issuing Protocols in ID-based Cryptosystems}
\author{Raju Gangishetti, M. Choudary Gorantla, Manik Lal Das, Ashutosh
Saxena\\
\\ Institute for Development and Research in Banking Technology\\
Castle Hills, Road \#1, Masab Tank, Hyderabad 500057, AP, INDIA. \\
\{graju, gmchoudary\}@mtech.idrbt.ac.in, \{mldas,
asaxena\}@idrbt.ac.in}  \maketitle
\begin{abstract}
To remove key escrow problem and avoid the need of secure channel
in ID based cryptosystem Lee et al.\cite{lb} proposed a secure key
issuing protocol. However we show that it suffers from
impersonation, insider attacks and incompetency of the key privacy
authorities. We also cryptanalyze Sui et al.'s\cite{sch} separable
and anonymous key issuing protocol.
\end{abstract}
\section{Review of Lee et al.'s Protocol \cite{lb}} It includes
five stages namely, System Setup, System Public Key Setup, Key
Issuing, Key Securing and Key Retrieving.

\subsection{System Setup} The KGC specifies two cyclic groups
$G_{1},G_{2}$ of prime order $q$ where $G_{1}$ is additive and
$G_{2}$ is multiplicative groups. It also defines a bilinear
mapping as $e:G_{1}\times G_{1} \rightarrow G_{2}$ between
$G_{1},G_{2}$ and hash functions $H:\{0,1\}^{*}\rightarrow G_{1}$,
$ h : G_{2} \rightarrow Z_{q}^{*} $. Let $P\in G_1$ be an
arbitrary generator of $G_1$. The KGC selects a master key $s_{0}
\in Z_{q}^{*}$ at random and computes its pubic key
$P_{0}=s_{0}P$.

\subsection{System Public Key Setup} The $n$ KPAs establish their
key pairs.  $KPA_{i}$ chooses his master key $s_{i}$ and computes
his public key $P_{i}=s_{i}P$,$\forall i=1,...,n$. Then all KPAs
cooperate sequentially and computes
$Y_{i}^{\prime}$=$s_{i}Y_{i-1}$ where
$Y_{0}^{\prime}=P_{0}=s_{0}P$.

Finally, $Y=Y_{n}^{\prime}=s_{0}s_{1}...s_{n}P$ is published as
system public key. This sequential process can be verified by
$e(Y_{i}^{\prime}, P)=e(Y_{i-1}^{\prime}, P_{i})$.

\subsection{Key Issuing} A user with $ID$ chooses a random secret
$x$, computes a blinding factor $X=xP$ and requests the KGC to
issue a partial private key by sending $X$, $ID$. Then the KGC
issues a blinded partial private key as follows.
\begin{enumerate}
    \item Checks the identification and computes the
    public key of the user as \\ $Q_{ID}=H(ID, KGC, KPA_{1},..., KPA_{n})$.
    \item Computes a blinded partial private key as
    $Q_{0}^{\prime}=h(e(s_{0}X,P_{0}))s_{0}Q_{ID}$.
    \item Computes KGC's signature on $Q_{0}^{\prime}$ as
    $Sig_{0}(Q_{0}^{\prime})=s_{0}Q_{0}^{\prime}$.
    \item Sends $Q_{0}^{\prime}$ and $Sig_{0}(Q_{0}^{\prime})$ to
    the user.
\end{enumerate}
The user can unblind $Q_{0}^{\prime}$ using his knowledge of $x$,
since\\ $h(e(s_{0}X,P_{0}))$=$h(e(s_{0}xP,P_{0}))$ =
$h(e(P_{0},P_{0})^{x}) $.

\subsection{Key Securing} The user requests $KPA_{i}(i=1,...,n)$
sequentially to provide key privacy service by sending $ID$, $X$,
$Q_{i-1}^{\prime}$ and $Sig_{i-1}(Q_{i-1}^{\prime})$. Then
$KPA_{i}$ performs following steps
\begin{enumerate}
    \item Checks
    $e(Sig_{i-1}(Q_{i-1}^{\prime}),P)=e(Q_{i-1}^{\prime},
    P_{i-1})$.
    \item Computes
    $Q_{i}^{\prime}=h(e(s_{i}X,P_{i}))s_{i}Q_{i-1}^{\prime}$ and
    $Sig_{i}(Q_{i}^{\prime}=s_{i}Q_{i}^{\prime})$.
    \item Sends $Q_{i}^{\prime}$ and $Sig_{i}(Q_{i}^{\prime})$ to
    the user.
\end{enumerate}

This process is carried out up to $KPA_{n}$. Finally user receives
$Q_{n}^{\prime}$.

\subsection{Key Retrieving} The user retrieves his private key
$S_{ID}$ by unblinding $Q_{n}^{\prime}$ as follows.
\begin{eqnarray*}
  S_{ID} &=& \frac{Q_{n}^{\prime}}
  {h(e(P_{0},P_{0})^{x})h(e(P_{1},P_{1})^{x})...h(e(P_{n},P_{n})^{x})}= s_{0}s_{1}...s_{n}Q_{ID}
\end{eqnarray*}
The user can verify the correctness of his private key by
$e(S_{ID},P)=e(Q_{ID},Y)$.

\section{Cryptanalysis of Lee et al.'s Protocol}
\subsection{Impersonation Attack } In Key Issuing phase, user sends
$X=xP$ and $ID$ to the KGC. Any active adversary can modify the
$X$ as $X^{*}=x^{*}P$ and still it cannot be detected by KGC.
Because there is no binding between the $ID$ and $X$. Then KGC
computes partial private key
$Q_{0}^{*}=h(e(s_{0}X^{*},P_{0}))s_{0}Q_{ID}$, and sends to the
user through public channel. Adversary can eavesdrop $Q_{0}^{*}$
and request the KPAs for key privacy service. At the end Adversary
can extract the private key by unblinding $Q_{n}^{*}$.

\subsection{Insider Attack}

In Key Securing phase, user requests $KPA_{i}$ to provide key
privacy service by sending $ID$, $X$, $Q_{i-1}^{\prime}$,
$Sig_{i-1}(Q_{i-1}^{\prime})$, where fourth parameter is a
signature of $KPA_{i-1}$ on third parameter.

If $KPA_{i-1}$ wants a signature of $KPA_{i}$ on $m$, he sends
$ID^{*}$, $X^{*}=x^{*}P$, $Q_{i-1}^{*}=rH(m)$  and
$Sig_{i-1}(Q_{i-1}^{*})=rs_{i-1}H(m)$ to $KPA_{i}$ where
$r\in_{R}Z_{q}^{*}$. Then $KPA_{i}$ performs the following steps

\begin{enumerate}
    \item Checks $e(Sig_{i-1}(Q_{i-1}^{*}),P) =
e(Q_{i-1}^{*}, P_{i-1})$.
    \item Computes
    $Q_{i}^{*}=h(e(s_{i}X^{*},P_{i}))s_{i}Q_{i-1}^{*}$ and
    $Sig_{i}(Q_{i}^{*})=s_{i}Q_{i}^{*}$.
    \item Sends $Q_{i}^{*}$ and $Sig_{i}(Q_{i}^{*})$ to the
    user(i.e. $KPA_{i-1}$).
\end{enumerate}

Now, $KPA_{i-1}$ has $Q_{i}^{*}=h(e(s_{i}X^{*},P_{i}))s_{i}rH(m)$
and he can extract the signature of $KPA_{i}$ on $m$ as
$h(e(P_{i},P_{i})^{x^{*}})^{-1} r^{-1} Q_{i}^{*}$ = $s_{i} H(m) $.
At the same time $KPA_{i}$ cannot get signature of the $KPA_{i-1}$
(i.e. $s_{i-1}H(m)$), because $KPA_{i-1}$ sends his signature in
blinded manner. Thus, $KPA_{i-1}$ can obtain $KPA_{i}$'s signature
on any message of his choice.

\subsection{Incompetency of KPAs} In Key Securing Phase, the user
requests $KPA_{i} (i=1,2,...,n)$ sequentially to provide key
privacy service by sending $ID$, $X$, $Q_{i-1}^{\prime}$, and
$Sig_{i-1}(Q_{i-1}^{\prime})$. Then $KPA_{i}$ validates the
received parameters by checking the equality\\
$e(Sig_{i-1}(Q_{i-1}^{\prime}),P)$=$e(Q_{i-1}^{\prime}, P_{i-1})$.

Any active adversary can alter $Q_{i-1}^{\prime}$,
$Sig_{i-1}(Q_{i-1}^{\prime})$ and replaces with the following
$Q_{i-1}^{*}=r^{*}Q_{i-1}^{\prime}$, $ Sig_{i-1}(Q_{i-1}^{*}) =
r^{*}Sig_{i-1}(Q_{i-1}^{\prime})$. Then $KPA_{i}$ performs
   \begin{enumerate}
        \item Checks $e(Sig_{i-1}(Q_{i-1}^{*}),P)=e(Q_{i-1}^{*},
    P_{i-1})$
        \item Computes
        $Q_{i}^{*}=h(e(s_{i}X,P_{i}))s_{i}Q_{i-1}^{\prime}$ and $ Sig_{i}(Q_{i}^{*}) =
s_{i}Q_{i}^{*}$
        \item Sends $Q_{i}^{*}$, and $Sig_{i}(Q_{i}^{*})$ to the user.
    \end{enumerate}

It may be noted that the user is not checking the correctness of
the received parameters in intermediate stages. Therefore any
modification by an Adversary during the communication between user
and $KPA_{i}$ will be undetected till the end of Key Securing
Phase. This requires the user to execute this phase again from the
beginning. Further, as the KGC and KPAs are not capable of
checking the validity of the received parameters, they are signing
them blindly.

The attack given in Section 2.1 can also be applied to
\cite{rcms}.
\section{Review of Sui et al. \cite{sch}} A one time password $pwd$ can
be established between the Local Registration Authority(LRA) and
the user after the off-line authentication. \\\textbf{Setup(run by
KGC)}: It takes the security parameter $k$ and returns $params$
(System Parameters) and the master-key. Let $G$ be a GDH group of
prime order $p$. Public information is $I_{SAKI}=(G,p,H,P_{PKG})$.
P is a generator of $G$ and $ H:{0,1}^{*}\rightarrow G$ is a
oneway hash function and $Q_{A}=H(id_{A})$. $P_{PKG}=sP$ is the
system public key.
\\\textbf{Key Generation}: It takes inputs as $params$,
master-key, and an arbitrary $ID\in \{0,1\}^{*}$; and returns a
private key $S_{ID}$. The password $pwd$ is user's chosen password
during off-line authentication and the tuple $(ID, pwd)$ is stored
in KGC's database of ``pending private key''.
\begin{enumerate}
    \item A:selects a random number $r$, $A \rightarrow
    KGC:Q=rH(ID), T=r^{-1}H(pwd)$.
    \item KGC: checks the validity of the request by checking
    whether $e(Q, T)=e(H(ID),H(pwd))$ holds for a certain tuple in
    KGC's database.
    \item KGC: computes $sQ$, $KGC \rightarrow A: S=sQ$
    \item A: verifies the blinded private key by checking
    $e(S,P)=e(Q, P_{PKG})$. If it holds, A unblinds the encrypted
    private key and obtains $sH(ID)$.
\end{enumerate}
The user can delete $pwd$ after obtaining the private key. The KGC
can also remove the tuple $(ID, pwd)$ from the database after the
protocol.

\section{Cryptanalysis of Sui et al. Protocol}

\subsection{Stolen Verifier Attack}
In Sui et al. protocol, ($ID$, password) is stored in KGC's
database. If an Adversary steals the database he can have genuine
users' secrets on requesting the KGC on behalf of any registered
user available in database. Though the KGC stores (ID, password)
for a short-time till the corresponding secret key is issued, it
affects the protocol entirely.

\subsection{Insider Attack} In practice, it is likely that a user
 uses same password to access several systems and other purposes
 for his convenience. In the registration phase, the user gives his
 password $pwd$ to LRA and the LRA stores the $ID$ and corresponding
 password in the database. In the extended scheme given to remove the
 key escrow by single KGC, the database is accessible by multiple KGC's
 and LRA. Any one of the insider of the system could impersonate user's
 login on stealing password and can get access of the other systems.

\subsection{Incompetency of KGCs} A user requests
for private key as follows:
\begin{itemize}
    \item Selects a random number $r$, and computes $Q=rH(ID),
    T=r^{-1}H(password)$ and sends to the KGC.
    \item KGC checks the validity of the request by checking the
    equality \\ $e(Q,T)=e(H(ID),H(password))$.
    \item Computes blinded private key $S=sQ$ and sends
    to the user where $s$ is the KGC's private key.
    \item Then user verifies $S$ by checking the equality
    $e(S,P)=e(Q, P_{pub})$ where $P_{pub}=sP$ is KGC's public key.
\end{itemize}

Any Adversary can alter the parameters $Q$, $T$  and replace with
$Q^{*}=r^{*}Q$, $T^{*}=r^{*^{-1}}T$ and KGC verifies the equality
$e(Q^{*},T^{*} )=e(H(ID),H(password))$. Then the KGC computes
$S^{*}=sQ^{*}$ and sends to the user. In this protocol  the KGC
cannot check the validity of the parameters received and thus
blindly signs on it.

\section{Conclusion}
In this work we have cryptanalyzed two ID based key issuing
protocols of \cite{lb, sch}. We showed that the Lee et al.
\cite{lb} protocol suffers from impersonation, insider attacks and
incompetency of the key privacy authorities. We also showed that
the Sui et al.'s\cite{sch} separable and anonymous key issuing
protocol suffers from stolen verifier, insider attacks and
incompetency of key generation centers.

\end{document}